\documentclass[notitlepage,twocolumn,superscriptaddress,amsmath,amssymb,aps,pra]{revtex4-2}
\usepackage{amsmath,amssymb,graphicx,mathtools,dsfont,ulem,bbold}
\usepackage[dvipsnames]{xcolor}
\usepackage[T1]{fontenc}
\usepackage{hyperref}
\usepackage{soul}
\usepackage{cleveref}
\usepackage{bibunits}
\hypersetup{citecolor=black,colorlinks=false,urlcolor=black}

\newcommand{\bra}[1]    {\langle #1|}
\newcommand{\ket}[1]    {|#1 \rangle}
\newcommand{\ketbra}[2]{|#1\rangle\!\langle#2|}

\usepackage{diagbox} 

\usepackage{hhline}

\begin{document}

\preprint{  }

\title{
Harnessing quantum back-action for time-series processing}

\author{Giacomo Franceschetto}
\email{giacomo.franceschetto@icfo.eu}
\affiliation{ICFO-Institut de Ciències Fotòniques, The Barcelona Institute of Science and Technology, 08860 Castelldefels (Barcelona), Spain}

\author{Marcin Płodzień}
\altaffiliation[Present address: ]{Qilimanjaro Quantum Tech, Carrer de Veneçuela 74, 08019 Barcelona, Spain }
\affiliation{ICFO-Institut de Ciències Fotòniques, The Barcelona Institute of Science and Technology, 08860 Castelldefels (Barcelona), Spain}

\author{Maciej Lewenstein}
\affiliation{ICFO-Institut de Ciències Fotòniques, The Barcelona Institute of Science and Technology, 08860 Castelldefels (Barcelona), Spain}
\affiliation{ICREA-Institució Catalana de Recerca i Estudis Avan\c cats, Lluís Companys 23, 08010 Barcelona, Spain}
 
\author{Antonio Acín}
\affiliation{ICFO-Institut de Ciències Fotòniques, The Barcelona Institute of Science and Technology, 08860 Castelldefels (Barcelona), Spain}
\affiliation{ICREA-Institució Catalana de Recerca i Estudis Avan\c cats, Lluís Companys 23, 08010 Barcelona, Spain}

\author{Pere Mujal}
\email{pere.mujal@alumni.icfo.eu}
\affiliation{ICFO-Institut de Ciències Fotòniques, The Barcelona Institute of Science and Technology, 08860 Castelldefels (Barcelona), Spain}

\date{\today} 
 
\begin{abstract}

Quantum measurements affect the state of the observed systems via back-action. While projective measurements extract maximal classical information, they drastically alter the system. %'s configuration. 
In contrast, indirect measurements balance information extraction with the degree of disturbance. Considering the prevalent use of projective measurements in quantum computing and communication protocols,  the potential benefits of indirect measurements in these fields 
remain largely unexplored. In this work, we show that measurement back-action, a purely quantum effect, can be harnessed within the quantum reservoir computing framework to enhance temporal data processing. Incorporating indirect measurements leads to improved execution-time scaling and overall performance. Our results reveal that carefully optimizing the measurement strength can significantly improve the quantum reservoir computing algorithm performance. Furthermore, our approach demonstrates improved memory performance when compared with state-of-the-art classical feedback protocols. This work provides a comprehensive and practical recipe to promote the implementation of indirect measurement-based protocols in quantum reservoir computing. Moreover, our findings motivate further exploration of experimental protocols that leverage the back-action effects of indirect measurements.

\end{abstract}

\maketitle

\section{Introduction}

A fundamental aspect that makes quantum measurements unique and counter-intuitive from a classical perspective is back-action, the influence of observers on the state of quantum systems. From a foundational perspective, this phenomenon is central to the ‘measurement problem', sparking formal and philosophical debates about different interpretations of quantum mechanics since von Neumann's time \cite{von1955mathematical, wheeler2014quantum}. From a more applied viewpoint, the so-called collapse of the wave function after the measurement process is crucial for different quantum technologies. For instance, in quantum key distribution \cite{BENNETT20147}, the disturbance caused by measurement is exploited to detect eavesdropping attempts, ensuring secure communication. Similarly, quantum sensing \cite{degen2017quantum} harnesses the extreme sensitivity of quantum systems to external interactions, achieving precision levels beyond classical methods. In this work, we explore a novel application of quantum back-action: enhancing the performance of machine-learning algorithms.

The degree of disturbance induced by a measurement depends on its nature. Conventional projective measurements cause substantial disruption to the quantum system, forcing it into the eigenstate (or a degenerate subspace) associated with the measured outcome. This strong interaction collapses superposition and fully determines the system’s state. In contrast, indirect measurements provide a gentler alternative by using indirect interaction between the system and the measurement apparatus \cite{10.1093/acprof:oso/9780199213900.001.0001}. The strength of the coupling and the duration of this interaction govern both the disturbance introduced by the measurement and the amount of information extracted. Consequently, indirect measurements can provide only partial information about the system, inducing a back-action that depends on the degree of measurement strength. This approach allows the quantum state to retain some coherence, enabling ongoing observation of the system without fully collapsing it. This makes indirect measurements a valuable tool for quantum information processing \cite{PhysRevA.62.012307, SINGH2014141, mujal2023time}.
Indirect measurements have been both demonstrated and widely used across various quantum platforms \cite{pan2020weak, spagnolo2022experimental,murch2013observing}. 
Superconducting qubits, for example, are typically measured through indirect measurements using ancillary microwave pulses \cite{naghiloo2019introduction,weber2016quantum}. By adjusting the strength and duration of these pulses, experimenters can tune the measurement strength from weak to strong, with the latter approaching projective measurements in the limit \cite{murch2013observing, Murch2016, doi:10.1126/science.1226897}.
In general, the capacity to implement indirect measurements relies on the ability to prepare convenient ancillas \cite{Yasuda2023} or to extract information through projectively measuring only a portion of the system \cite{hu2024overcoming, viqueira2023density}.
\begin{figure*}[ht!]
\centering
\includegraphics[width = \textwidth]{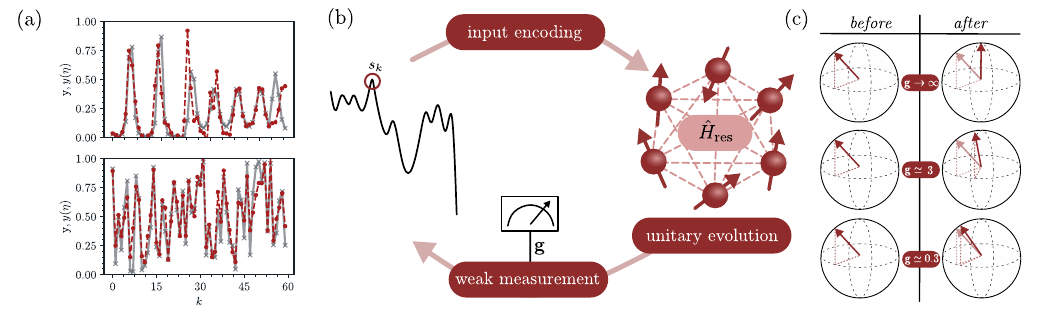}
\caption{Quantum reservoir computing via indirect measurements. (a) Example of the reservoir's prediction (dashed lines) over the target series (solid lines) on two specific instances ($\eta = 5$) for the forward prediction task (top panel) and memory retrieval task (bottom panel) (see section \ref{sectasks}). (b) Schematic view of the information processing step of the QRC algorithm, the coupling strength $g$ between the measurement apparatus and the reservoir is the main parameter of this work's study. (c) Illustrative example of the effect of indirect measurement on a single spin state for different values of the measurement strength, $g$. As $g \to \infty$, the measurement becomes equivalent to a projective measurement, fully collapsing the state onto one of the measurement axes. However, as $g$ decreases, the post-measurement state is no longer collapsed along the measurement axis and is less perturbed. In this analysis, we assume measurements are performed along the $z$-axis.  } \label{fig:widefig}
\end{figure*}

We illustrate our findings within the context of time-series processing using quantum reservoir computing (QRC) \cite{mujal2021opportunities}. QRC belongs to the broader category of quantum machine learning, a field that emerged from the desire to combine machine learning techniques with the unique advantages of quantum systems \cite{dawid2023}. First introduced in \cite{fujii2017harnessing}, it is a supervised quantum machine learning technique that leverages the advantages of quantum systems as the exponential growth of their state space with increasing system size. At the same time, it incorporates the simple and efficient training strategy characteristic of classical reservoir computing \cite{jaeger2004harnessing}, where the internal dynamics of the reservoir are fixed and only the output layer is trained. Recently, there has been growing interest in QRC and related approaches, with progress both in theoretical advancements \cite{Nakajima2019boosting,Ghosh2019,Ghosh2019prl,Kutvonen2020,martinez2020information,Ghosh2020reconstructing,Tran2020higherorder,Nokkala2021,martinez2021dynamical,Ghosh2021neuromorphic,mujal2021analytical,govia2022nonlinear,Angelatos2021qstatemeas,Nokkala2022,Llodra2022,Domingo2022,Yelin2022,Nokkala2023,GarciaBeni2023,martinez2023finitedim,Krisnanda2023tomographic,mujal2023time,Innocenti2023,Wudarski2023,GarciaBeni2024squeezing,sannia2024dissipation,Nokkala2024,Jaurigue2024memory,Li2024estimating,Kobayashi2024,Domingo2024,Kobayashi2024eco,palacios2024role,Llodra2024} and experimental implementations on diverse platforms \cite{negoro2018machine,Chen2020,Suzuki2022,spagnolo2022experimental,Pfeffer2022,Kubota2023,Yasuda2023,hu2024overcoming,Paternostro2024,Yelin2024}.

In QRC, memory refers to the reservoir's ability to retain and utilize information from past inputs to inform future predictions. 
It plays a key role in processing temporal series by capturing the time correlations within the dataset. The amount of memory retained by quantum reservoirs has been shown to depend on their dynamical regime \cite{martinez2021dynamical} and to be reduced by the effects of quantum back-action \cite{mujal2023time}.
When online processing time series, projective measurements introduce significant decoherence, which negatively impacts the memory capacity of the reservoir. As a result, to truly take advantage of the scaling benefits in online processing, a weak approach that minimizes the system disturbance is more suitable \cite{mujal2023time}. In this case, there is a tradeoff between the amount of information extracted and the quantum back-action introduced. Although reducing the disturbance preserves the reservoir’s memory, it also limits the information available, leaving the potential advantage uncertain. This approach has already attracted experimental interest and has been realized in superconducting devices through mid-circuit measurements \cite{hu2024overcoming,Yasuda2023}.

In this work, we reveal that adjusting the strength of quantum measurements enables the optimization of quantum back-action introduced into the system’s evolution, thereby enhancing the performance of machine learning tasks beyond what is achievable without this control.
Interestingly, in various contexts, controlling the degree of decoherence introduced into the reservoir has been shown to improve its memory capacity \cite{martinez2023finitedim, sannia2024dissipation, monzani2024leveraging,palacios2024role}. In this work, we build on this insight by systematically exploring the entire range of indirect measurement regimes and jointly optimizing the measurement strength with the reservoir Hamiltonian parameters, in order to clarify its fundamental role in the reservoir’s performance. By leveraging the ability to tune the measurement strength, we optimize the tradeoff between information extraction and back-action. This allows for an online processing scheme that benefits from linear scaling without suffering drawbacks from decoherence. Instead, by carefully controlling the measurement strength, we can enhance the reservoir’s capacity. To achieve this, we introduce a new figure of merit to compare the performance of the indirect measurement-based protocol, where back-action is controlled, with projective measurement-based protocols that rely on restarting strategies to avoid back-action. By optimizing this metric, we identify the optimal measurement strength that maximizes performance. We illustrate this improvement across different realizations of reservoirs operating in distinct dynamical regimes, validated through two benchmark tasks: the forecasting of the Santa Fe chaotic series \cite{PhysRevA.40.6354} and the short-term memory task . Furthermore, we extend our analysis to a realistic finite resource scenario and show that the enhancement effect is amplified by the favourable scaling properties of the indirect measurement protocol. Finally, we compare our approach with a classical feedback strategy, demonstrating the advantage of exploiting a pure quantum effect such as back-action with respect to incoherent feedback.

\section{Preliminaries}
\subsection{Quantum reservoir computing via indirect measurements}
Reservoir computing was proposed as a training-efficient alternative to more complicated recurrent neural network models like Long Short-Term Memory for processing temporal data and time series \cite{jaeger2004harnessing}. Quantum reservoir computing extends the concept by leveraging quantum properties and dynamics, using a quantum system as the reservoir \cite{fujii2017harnessing}. The general pipeline of QRC algorithms is structured in two steps: first the reservoir has to process the time series of interest and then a linear regression model is trained to perform predictions (Fig. \ref{fig:widefig} (a)). While the latter is a straightforward process, the former includes three stages to be repeated for each element of the dataset: input encoding, time evolution and information extraction (Fig. \ref{fig:widefig} (b)). At the encoding stage, the information of the $k$-th element of the time series is injected into the amplitudes of the state of a fixed spin of the reservoir \cite{fujii2017harnessing, martinez2021dynamical, mujal2021analytical}. The information is then spread through the system by evolving it for a time $\Delta t$ under a transverse field Ising Hamiltonian $\hat{H}$. 
Ultimately, we measure a set of observables $\{\langle \hat{O}_i \rangle\}$ to extract information from the quantum system. However, when performing projective measurements, it is convenient to reset the system and reprocess the previous $k$ elements without applying measurements before processing the next input, $k+1$. This is because projective measurements erase the stored information in quantum coherences through a destructive process. This procedure, known as the restarting protocol (RSP), has long served as the canonical benchmark in the field, since it introduces no additional disturbance to the system.

Beyond full restarting, an alternative strategy is the rewinding protocol (RWP)~\cite{Chen2020,mujal2023time}. Instead of repeating the entire past evolution after every projective measurement, the RWP leverages the fading-memory property of the reservoir and rewinds only the last $\tau_{\mathrm{wo}}$ steps. Concretely, at each output extraction the system is re-prepared at time $t_k-\tau_{\mathrm{wo}}$ and only the corresponding $N_{\mathrm{wo}}=\tau_{\mathrm{wo}}/\Delta t$ inputs are re-injected before measurement (see Fig.~1(c) in Ref.~\cite{mujal2023time}). This reduces the repetition overhead from quadratic to linear in the dataset size while still preventing the accumulation of measurement back-action. Still, the RWP relies on projective measurements and explicit replay of past inputs, making it intrinsically non-online. This has motivated alternative approaches that preserve single forward-pass reservoir evolution while achieving competitive performance \cite{settino, hu2024overcoming, Kobayashi2024}, including indirect measurement–based schemes introduced in Ref. \cite{mujal2023time}. In this scheme, referred to as the online protocol (OLP), only a small amount of information is lost due to the back-action effect (Fig. \ref{fig:widefig} (c)), allowing the QRC algorithm to run continuously without restarting at each step in the time series processing. The OLP introduces back-action during the evolution as a tradeoff, allowing for better algorithm scaling by avoiding the need to restart at each input.

\subsection{Two benchmark studies}
\label{sectasks}
We perform an in-depth analysis of how the back-action effect due to using an indirect measurement-based protocol affects the performance of QRC. To do so, we take into account two tasks: the forecasting of a chaotic time series and the efficiency in storing past information. The chaotic time series we use in this work is a subset of $K=2000$ elements of the Santa Fe laser time series, a common benchmark dataset that follows the chaotic behavior of a $NH_3$ laser \cite{PhysRevA.40.6354}, the exact data points can be found in Ref. \cite{git_tuto}. On the other hand, to test the ability to retrieve past information through the short-term memory task, we generate a set of $K=$ 1000 random points from the uniform distribution in the range $[0, 1]$. We build $\eta_\text{max}$ different sub-tasks for both benchmarks where $\eta$, the prediction distance, varies from $1$ to $\eta_\text{max}$. Given an $\eta$, the target series $k$-th element becomes $t_{k}(\eta) = s_{k \pm \eta}$, where $\vec{s}$ is the original time series. The capacity of the reservoir for each sub-task is evaluated by computing the coefficient of determination $C(\eta)$ of the algorithm predicted series $\vec{p}$ and the sub-task corresponding true one $\vec{t}(\eta)$:
\begin{equation}
    C(\eta) = \frac{\text{cov}^2(\vec{p}, \vec{t}(\eta))}{\text{var}(\vec{p})\text{var}(\vec{t}(\eta))}.
\end{equation}

The predictions of the QRC algorithm are generated by a linear model trained on $70\%$ of the output dataset, which consists of the expectation values of the measured observables for each time series element. An example of the predictions of the QRC algorithm on a specific instance of the two tasks is presented in Figure \ref{fig:widefig} (a).
Then, the overall performance of the reservoir on the task is measured through the sum capacity:
\begin{equation}
    C_{\Sigma} = \sum_{\eta = 1} ^{\eta_\text{max}} C(\eta)\label{eq:sumcap},
\end{equation}
where $\eta_\text{max}$ denotes the $\eta$ for which $C(\eta)\approx0$.
\section{Methods}
\subsection{Reservoir Hamiltonian}
The quantum reservoir considered in this work consists of a system of $N$ spins with an all-to-all connection scheme, where the dynamics is governed by the transverse-field Ising Hamiltonian:
\begin{equation} \label{eq:res_ham}
    \hat{H} = \sum _{i<j=0} ^{N-1} J_{ij} \hat{\sigma}^x_i\hat{\sigma}^x_j + \frac{h}{2} \sum_{i=0} ^{N-1} \hat{\sigma}^z_i.
\end{equation}

In the numerical simulations, we considered systems of $N=6$ spins with random couplings $J_{ij}$ uniformly sampled from the range $[-J/2, J/2]$ and we set $J = 1$. Then, when referring to the other parameters of the Hamiltonian we always implicitly express them in units of $J$.
At the beginning of the protocol, the initial state of the system is set to be $\ket{\Psi} = \ket{0} ^{\otimes N }$.

\subsection{ Details of the algorithm pipeline}
Focusing on the first step of the QRC algorithm pipeline, we provide a detailed description of the framework used in this work. To process the time series, firstly is necessary to normalize it and encode each element in the quantum system. Among the $N$ spins of the reservoir, we consider the following bipartition: we fix one spin to act as the input node, partition $A$, while the others $N-1$ form the partition $B$. For each element of the sequence $s_k$, the encoding routine consists of resetting $\hat{\rho}^A$ to the pure state $\ket{\psi_k} = \sqrt{1-s_k}\ket{0}+\sqrt{s_k}\ket{1}$ \cite{mujal2021analytical}. After the input injection, the reservoir evolves under $\hat{H}$ for a time $\Delta t=10/J$, $\hat{U}=e^{-i\hat{H}\Delta t}$. The duration of $\Delta t$ is chosen following Ref.~\cite{martinez2020information}, ensuring that the reservoir exhibits a sufficiently rich dynamical response capable of supporting both linear and nonlinear memory.
In summary, if before the encoding of the element $s_{k}$ the state of the reservoir was $\hat{\rho}_{k-1}$, afterward the reservoir undergoes a completely positive trace-preserving (CPTP) map \cite{fujii2017harnessing, martinez2021dynamical}:
\begin{equation}
    \mathcal{L}_k (\hat{\rho}_{k-1}) = \hat{U} (\ketbra{\psi_k}{\psi_k} \otimes {\rm Tr_A}(\hat{\rho}_{k-1}))\hat{U}
^\dagger.
\end{equation}

At this point, before the next input injection, it is necessary to extract the elaborated information from the system. As previously described, here we use the online protocol proposed in \cite{mujal2023time} based on indirect measurements.
These are realized by measuring the momentum quadrature of an ancillary coherent-state pulse coupled to the indirectly measured qubit. As a result, the following measurement operator, e.g. along the $z$-direction, is implemented:

\begin{equation}\label{eq:operValso}
    \hat{\Omega}^{z}_V=\frac{1}{\sqrt[4]{2\pi}}\left(e^{-\frac{(V-g)^2}{4}}\ket{0}\bra{0}+e^{-\frac{(V+g)^2}{4}}\ket{1}\bra{1}\right),
\end{equation}
where the continuous-valued measurement outcome $V$ and the measurement strength $g$ are in units of the Gaussian width (see Supplementary Information of \cite{mujal2023time} for details).
The effect of this measurement on the $z$-axis direction on the unconditional state of the reservoir can be modeled through the action of the matrix $\hat{M}=(\mathbf I + e^{-\frac{g^2}{2}}\hat{\sigma}^x)^{\otimes N }$, such that the state after a measurement with strength $g$ is given by:
\begin{equation}
    \hat{\rho}_k = \hat{M} \odot {\mathcal{L}}_k (\hat{\rho}_{k-1}),
\end{equation} 
where $\odot$ indicates the element-wise product.
Actually, the CPTP map in this last equation is equivalent to a dephasing channel \cite{martinez2023finitedim}.
We can then extract the expectation value of the observable $\hat{\sigma}_i^z$ on the i-th spin as ${\rm Tr}(\hat{\rho}_k \hat{\sigma}_i^z)$. To measure single qubit observables in the $x$ and $y$ direction it is necessary to apply proper rotations to ${\mathcal{L}}_k (\hat{\rho}_{k-1})$ before applying $\hat{M}$. For two-spin observables the same description applies. Since the RSP is equivalent to dynamics without back-action, it can be conveniently simulated by setting $g=0$ in this model, for which $M$ reduces to an all-ones matrix and has no effect on the state. For our purposes, the RWP is simulated identically to the RSP. Because rewinding reproduces the same unperturbed reservoir dynamics after the short washout transient, the ideal RWP coincides with the RSP at the level of state evolution. Their difference manifests only when accounting for finite resources, where the two protocols incur different repetition overheads, but this affects only the practical resource cost and not the simulated dynamics themselves.

After having processed all the elements of the sequence $\vec{s}$, we obtain a dataset where for each element of the input data vector we have the corresponding expectation values of the single-spin observables
$
\langle \hat{\sigma}^x_i \rangle,  
\langle \hat{\sigma}^y_i \rangle,
\langle \hat{\sigma}^z_i \rangle
$
for each spin in the reservoir, as well as the two-spin observables
$\langle \hat{\sigma}^x_i \hat{\sigma}^x_j \rangle, 
\langle \hat{\sigma}^y_i \hat{\sigma}^y_j \rangle, 
\langle \hat{\sigma}^z_i \hat{\sigma}^z_j \rangle $
for all pairs of spins \(i \neq j\). Single-spin variances, i.e.,
$\langle (\hat{\sigma}^\alpha_i)^2 \rangle,$
are also computed as a consistency check, though they do not affect the performance. Since measuring in the $x$ and $y$ direction implies applying rotations, this means that the algorithm has to go through the entire dataset three times and not just once. 

From the obtained $(K \times 81)$ dataset, we discard the first $20$ rows, which correspond to the expectation values of the initial $20$ elements of the time-series.
This washout step and the fading-memory property of the reservoir ensure that its state becomes independent of the initial conditions and depends only on the recent input history.
The choice of $20$ steps follows the analysis of the required washout length for this system reported in Ref.~\cite{martinez2020information}.

\section{Results}
\subsection{Performance analysis in the infinite resources scenario}
Previous work~\cite{mujal2023time} revealed that the same performance achieved by protocols following the unperturbed reservoir dynamics, such as the RSP and the RWP, can, in some cases, also be obtained with the OLP, despite the introduction of back-action effects.
 On the basis of this, we analyze the performance of QRC with the OLP at different measurement strengths. Moreover, we consider different external field strengths, $h$, i.e. different dynamical regimes of the reservoir \cite{martinez2021dynamical}. Both quantities are expressed as dimensionless throughout this work.

To quantify the performance of the OLP with respect to protocols that preserve the unperturbed reservoir dynamics, we introduce the performance ratio $P_R$. Since the RSP and the RWP are equivalent in the ideal limit, as both reproduce the same unperturbed trajectory, we use the RSP as representative of the unperturbed dynamics. In its simplest form, $P_R$ compares the OLP at fixed $(g,h)$ with the unperturbed dynamics evaluated at the same field value $h$:
\begin{equation}
    P_R(g,h) = \frac{C_{\Sigma}^{\text{OLP}}(g,h)}{C_{\Sigma}^{\text{RSP}}(h)}.
\end{equation}
This is the key quantity for our analysis, as it directly reveals whether measurement back-action helps or hinders the performance of the reservoir for that particular configuration.

For completeness, we also consider a stricter performance indicator, $P_R^{\min}$, where the denominator is replaced by the best-performing unperturbed reservoir across all scanned values of~$h$:
\begin{equation}
    P_R^{\min}(g,h) = \frac{C_{\Sigma}^{\text{OLP}}(g,h)}{\max_{h'}\, C_{\Sigma}^{\text{RSP}}(h')}.
\end{equation}
While $P_R$ captures the role of back-action for the given reservoir, $P_R^{\min}$ answers the stronger question of whether the OLP can outperform even the optimal unperturbed configuration. Because the benchmark in the denominator is the best reservoir available, this metric is intentionally strict: values $P_R^{\min}>1$ are attained only when the reservoir is intrinsically very good and, in addition, back-action provides a genuine performance enhancement.
 One could also choose to maximize $P_R$ or $P_R^{\min}$ separately for each direction, thereby obtaining optimal parameters independently along each axis. In Appendix B, we show that in our case this does not improve the results; however, it could guide an informed choice of the preferred measurement direction when resources are limited or bring advantages if one cares about a specific task with small $\eta$.

We average the $P_R$ over 50 different reservoir realizations, where the couplings in between the spins are randomly selected each time. In the subsequent plots, the lines indicate the mean values, while the shaded regions denote one standard deviation around the mean.
\begin{figure}[t!]
\includegraphics[width = 0.48\textwidth]{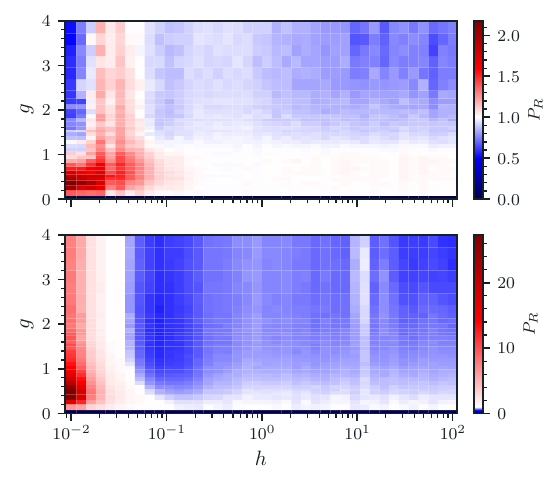}
\caption{Performance ratio analysis for the forward prediction task (top) and for the short-term memory task (bottom). A $P_R$ in the range $[0, 1)$ shows that the unperturbed dynamics protocol is performing better than the OLP in that point of the $g$--$h$ map. A $P_R$ around 1 shows that the OLP is performing as good as the RSP, and over 1 indicates an over-performance due to the effect of back-action. The maximum $P_R$ and corresponding parameter values for each task are: forward prediction: $g^* = 0.355$, $h^* = 0.01$, $P_R = 2.178$; short-term memory: $g^* = 0.486$, $h^* = 0.01$, $P_R = 27.302$.}
 \label{fig:cmaps_pr}
\end{figure}
In Figure~\ref{fig:cmaps_pr} we show how $P_R$ varies across different values of $g$ and $h$ for the forward prediction task on the Santa Fe dataset (top panel), and for the short-term memory task (bottom panel). Each colormap is composed of 33 by 40 points and is computed under the assumption of having access to an infinite number of shots for each measured observable. Regions where $P_R>1$ indicate that back-action provides an advantage over the corresponding unperturbed reservoir. The $P_R$ at $g=0$ is set to be $0$ since it corresponds to the case in which the measurement apparatus is not interacting with the reservoir and the measurement output is then a random number uncorrelated to the system.

A consistent and evident performance enhancement is observed across broad regions of the parameter space. In particular, back-action is most beneficial when the unperturbed dynamics struggles the most, such as at very small values of $h$, where the reservoir becomes fully localized and the RSP performance drops significantly. In these regimes, the OLP can substantially improve the effective dynamics, reaching enhancement factors of up to $P_R \approx 2$ for the Santa Fe task and $P_R \approx 27$ for the short-term memory task. When the unperturbed dynamics already performs well, the values of $P_R$ typically remain close to one, showing that the OLP can match the performance of the unperturbed reservoir. As we discuss in the next analysis based on $P_R^{\min}$, identifying operating points where back-action also improves upon the best unperturbed configuration requires a more stringent comparison.

\begin{figure}[t!]
\includegraphics[width = 0.48\textwidth]{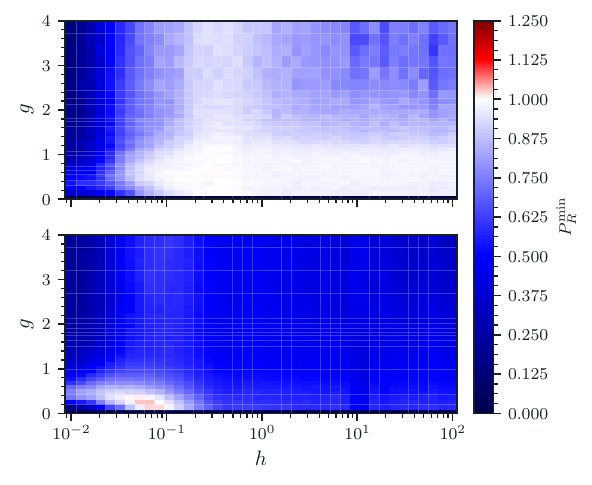}
\caption{Minimum performance ratio analysis for the forward prediction task (top) and for the short-term memory task (bottom). A $P_R^{\min}$ in the range $[0, 1)$ shows that the best performing RSP is performing better than the OLP in that point of the $g-h$ map. A $P_R^{\min}$ around 1 shows that the OLP is performing as good as the best performing RSP, and over 1 indicates an over-performance. The optimal parameters and corresponding $P_R^{\min}$ values for each task are: forward prediction — RSP: $h'^* = 0.346$; OLP: $g^* = 0.355$, $h^* = 0.273$, $P_R^{\min} = 1.006$; short-term memory — RSP: $h'^* = 0.084$; OLP $g^* = 0.256$ , $h^* = 0.066$ , $P_R^{\min} = 1.029$.} \label{fig:cmaps_prmin}
\end{figure}
In Figure~\ref{fig:cmaps_prmin} we show the behaviour of $P_R^{\min}$ across the same $(g,h)$ grid used in the previous colormaps (33 by 40 points, infinite-shot assumption) for the forward prediction task on the Santa Fe dataset (top) and for the short-term memory task (bottom). Since $P_R^{\min}$ compares the OLP with the best-performing RSP over all scanned values of $h$, it provides a stricter benchmark.

For both tasks, a region with $P_R^{\min}>1$ clearly emerges, showing that back-action plays a role even for those OLP configurations that yield the highest capacities. The effect is smaller than in the $P_R$ maps, but remains visible: for the Santa Fe task the advantage concentrates at small $h$, where the unperturbed dynamics is most limited, while for the short-term memory task a narrower enhancement region appears around $h \sim 10^{-1}$, consistent with the location of the optimal unperturbed performance.

Overall, the two sets of colormaps convey a consistent picture. The $P_R$ maps show that back-action can substantially enhance the performance of the reservoir, especially in regions where the unperturbed dynamics is most limited. In these regimes the OLP not only compensates for the information extracted during the indirect measurement, but can even surpass the corresponding unperturbed dynamics. The $P_R^{\min}$ maps confirm that this positive contribution persists also for the best-performing OLP configurations, although the effect is naturally smaller due to the stricter benchmark. Together, these observations demonstrate that a properly tuned indirect-measurement protocol can retain the full capabilities of the unperturbed dynamics while also providing genuine performance gains in suitable regions of parameter space.

\begin{figure}[t!]
\includegraphics[width = 0.48\textwidth]{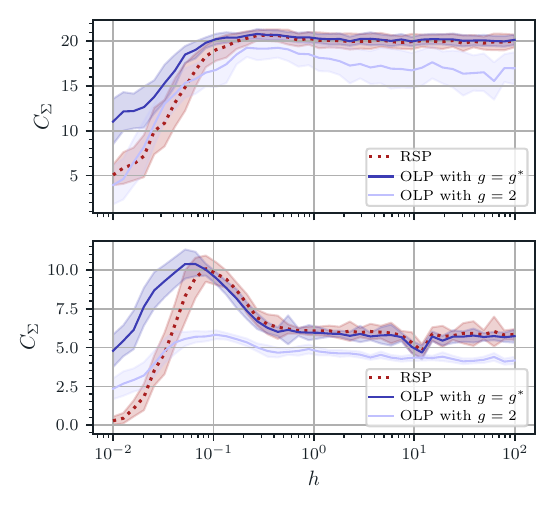}
\caption{ Sum capacity comparison of the RSP with respect to the OLP with and without optimizing the strength of the measurement $g$. With $g^*$ we denote the $g$ value that maximizes the $P_R^{\min}$. The top panel shows the results for the forward prediction task while the bottom panel is for the short-term memory task.
}\label{fig:cuts}
\end{figure}

To stress this last point, in Figure \ref{fig:cuts}, we compare the different sum capacities of the QRC running with the RSP and with the OLP in the case in which $g$ is fixed to a non-optimal value and in the case in which it is fixed to the value that maximized $P_R^{\min}$ in Figure \ref{fig:cmaps_prmin}. As a first observation, we can refer to \cite{martinez2021dynamical} to justify the dependence of the performance on the $h$ value. In fact, along the horizontal axis, the quantum system undergoes a transition from a localized regime to an ergodic one and it was showcased that being in the ergodic regime is a preferential condition to perform reservoir computing. Moreover, what emerges clearly is that optimizing the measurement strength, rather than simply minimizing back-action, is crucial for improving the protocol.

\begin{figure}[]
\includegraphics[width = 0.48\textwidth]{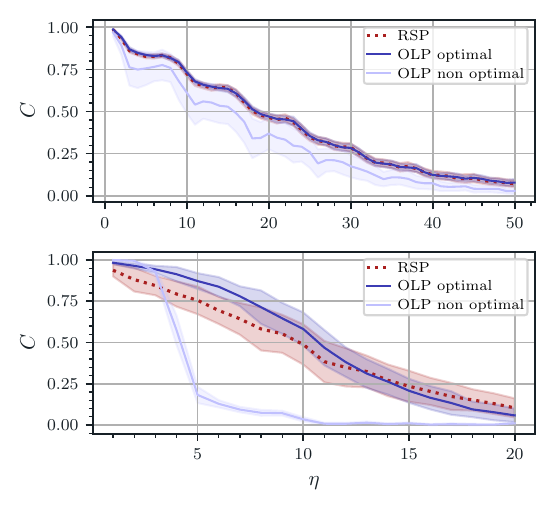}
\caption{Capacity for different sub-tasks. The figure compares the performance in single sub-tasks of the optimal restarting protocol with respect to the optimal and non-optimal versions of the OLP. The top panel shows the results for the forward prediction task while the bottom panel is for the short-term memory task. The optimal parameters correspond to the values obtained in Figure \ref{fig:cmaps_prmin}, whereas the non-optimal parameters are fixed to $g = 2$ and $h = 38.88$.
} \label{fig:cuts_c}
\end{figure}

The previous analysis is related to the sum capacity of the QRC across two different tasks. However, for different sub-tasks, defined by varying $\eta$, the performance may systematically differ. In this context, Figure \ref{fig:cuts_c} compares two points in the $g-h$ parameter space explored in Fig. \ref{fig:cmaps_prmin}, the one that maximized $P_R^{\min}$ and a non-optimal point. We then evaluate the performance for each different sub-task relative to the performance of the RSP at optimal $h$.
For the forward prediction task, we observe that, given an optimal external field and measurement strength, the OLP consistently outperforms the RSP for every value of $\eta$. In contrast, for the memory retrieval task, the optimal OLP outperforms the RSP only for $\eta$ values less than 10, while the RSP performs better for the other sub-tasks.
Moreover, in the upper panel, a non-optimal choice of parameters results in a constant factor decrease in performance. Instead, for the short-term memory task, the choice of parameters significantly impacts performance across almost all sub-tasks. Therefore, depending on the specific task, selecting the appropriate values for $g$ and $h$ is crucial when using the OLP, as illustrated by the performance at high $\eta$ with non-optimal parameters in the bottom panel of Fig. \ref{fig:cuts_c}.

\subsection{Considering a finite amount of resources}
In the previous section, we analyzed the effect of making use of an indirect measurement-based approach to QRC in the ideal scenario of infinite resources. Here, we consider the case in which each expectation value extracted from the reservoir is computed with a finite number of shots $N_s$. To take this into account, it has been shown that it is enough to add a Gaussian noise dependent on $N_s$ and $g$ to the previously computed expectation values \cite{mujal2023time}. 
Namely, for one and two spins observables the following statistical uncertainties are considered:
\begin{equation}\label{eqsuncertainties}
    \overline{s}_{\hat{\sigma}} = \sqrt{\frac{g^2+1}{g^2N_s}}, \quad  \overline{s}_{\hat{\sigma}\otimes\hat{\sigma}} = \sqrt{\frac{g^4+2g^2+1}{g^4N_s}}.
\end{equation}

For the RSP, one simply considers the asymptotic regime of very large $g$.
In this context, it is important to consider the inherent inefficiency of the RSP compared to the OLP, as discussed in \cite{mujal2023time}. This analysis takes into account the time resources required for executing each protocol. With the RSP, the time required to run the QRC scales quadratically with respect to the dimension of the time series, whereas for the OLP, this dependence is linear. Therefore, by fixing the number of shots available for the OLP, we can compute the required time and subsequently determine the number of shots available for the RSP within that same time frame. Approximately, we find that the number of shots available for the RSP is proportional to the number available for the OLP divided by $K$, the size of the dataset (see Appendix A). Finally, for completeness, we also consider the rewinding protocol (RWP). Since the RWP shares the same linear scaling in time as the OLP, both protocols can be allocated the same number of shots for a fair comparison at finite resources. This makes the RWP a natural unperturbed-dynamics benchmark in the finite-resource setting, allowing us to isolate the contribution of measurement back-action without the additional overhead present in the RSP.

As a first step, we focus on the practical and strict performance indicator $P_R^{\min}$ applied to the RSP. The goal is to quantify, in a realistic setting with finite resources, the combined effect of measurement back-action and the intrinsic scaling limitations of the RSP. Since the RSP becomes increasingly inefficient as the length of the time series grows, it provides a relevant baseline for evaluating how back-action interacts with sampling noise and resource constraints. 

\begin{figure}[t!]
\includegraphics[width = 0.48\textwidth]{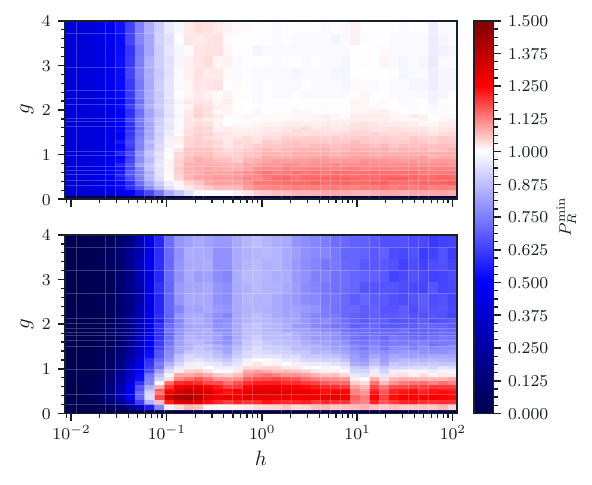}
\caption{Minimum performance ratio analysis for the forward prediction task (top) and for the short-term memory task (bottom) in the scenario of finite resources. We fixed $N_s^{\text{OLP}} = 1.5 \cdot 10^6$ and computed $N_s^{\text{RSP}}$ consequently for both tasks ($\approx 1.5 \cdot 10^3$ for the forward prediction task and $\approx 3 \cdot 10^3$ for the short-term memory task). The optimal parameters and corresponding $P_R^{\min}$ values for each task are: forward prediction — RSP: $h'^* = 15.118$; OLP: $g^* = 0.486$, $h^* = 30.703$, $P_R^{\min} = 1.158$; short-term memory — RSP: $h'^* = 1.245$; OLP $g^* = 0.355$ , $h^* = 0.170$ , $P_R^{\min} = 1.416$.}\label{fig:cmaps_real}
\end{figure}

Using $P_R^{\min}$ is particularly appropriate in this context: by comparing the OLP with the best-performing unperturbed reservoir across all scanned parameters, this measure allows us to visualize the behaviour precisely in the regime where the reservoir performs well and where meaningful improvements are most significant. In other words, $P_R^{\min}$ highlights whether back-action can still provide an advantage even when the unperturbed dynamics is operating close to its optimal configuration, making it a stringent and practically motivated first benchmark before moving to more efficient protocols such as the RWP.
We therefore compute $P_R^{\min}$ for the RSP (Fig.~\ref{fig:cmaps_real}). In this analysis, we fix $N_s^{\text{OLP}} = 1.5 \cdot 10^6$. As before, $P_R^{\min}$ is evaluated on a $33 \times 40$ grid in the $g$--$h$ parameter space, and each point corresponds to an average over 50 different reservoir realizations.

In this case, the same clear message emerges from the analysis of $P_R^{\min}$ in both tasks: in the finite-resources scenario defined above, using the OLP with properly tuned $g$ and $h$ leads to higher performance with respect to the unperturbed dynamics of the RSP, which is intrinsically inefficient at finite resources. This conclusion is conveyed by the red regions that appear in the plots for $h \ge 10^{-1}$ and $g \le 1$. In these regions, $P_R^{\min}$ is substantially greater than $1$, meaning that the OLP is performing better than the best-performing RSP under the same resource budget. As discussed earlier, this behaviour reflects the combined effect of the higher efficiency of the OLP at finite resources and the positive contribution of back-action in the regime where the measurement strength is optimally tuned.

While this analysis is very useful in practice, it is difficult to isolate the true contribution of back-action when comparing the OLP with the RSP, since the finite-resource inefficiency of the RSP also plays an important role. For this reason, we complement the previous study with a comparison against the RWP. The RWP provides an efficient unperturbed-dynamics benchmark, as it shares the same linear scaling as the OLP; both protocols can therefore be allocated the same number of resources. This allows us to disentangle the effects of sampling noise from those of measurement back-action.

\begin{figure}[t!]
    \centering
    \includegraphics[width=0.48\textwidth]{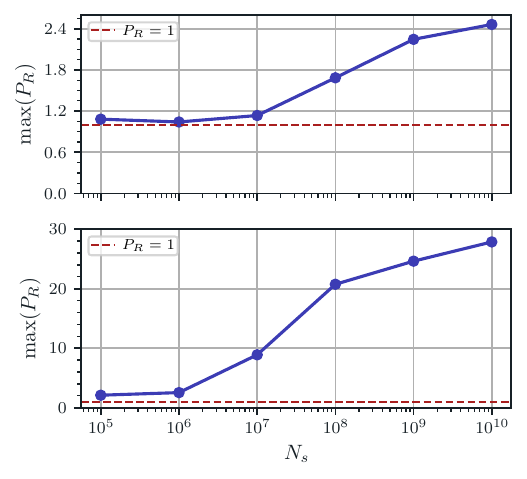}
    \caption{Maximum value of $P_R$ when comparing the OLP with the RWP, as a function of the total number of measurements $N_s$, for the forward prediction task (top) and for the short-term memory task (bottom). The red dashed line marks the threshold $P_R = 1$. For both tasks, the maximal $P_R$ remains above one even at modest sampling, indicating that back-action provides a positive contribution to the performance of the OLP. As $N_s$ increases, the maximal $P_R$ grows and approaches the ideal values obtained in the infinite-resources scenario.}
    \label{fig:maxPR_vs_resources}
\end{figure}

In this setting, we consider the maximal value of $P_R(g,h)$, now defined with respect to the RWP, ensuring a fair comparison between protocols with identical resource budgets. As shown in Fig.~\ref{fig:maxPR_vs_resources}, we compute this quantity for different total numbers of resources and observe that the contribution of back-action remains significant even at finite sampling. Increasing the number of resources improves the performance, as the OLP requires higher sampling resolution to fully exploit the expectation values of its observables. This behaviour shows that the advantage observed in the ideal scenario is not an artifact of assuming infinite resources, but persists in practical conditions and progressively approaches the ideal limit as more measurements are available.

\subsection{Comparison with a classical feedback-based protocol}
So far we have compared the OLP driven by an optimal back-action with the restarting protocol. This allowed us to highlight the advantage of exploiting the information disturbance tradeoff with respect to an ideal but inefficient scenario were measurements do not affect the system. More recently, alternative online schemes have been proposed where the reservoir is restarted at every step but complemented with a classical feedback layer that reconstructs the reservoir state informed by the previous expectation values \cite{Kobayashi2024, monomi2025feedbackenhancedquantumreservoircomputing}. This approach is undoubtedly practical and efficient, yet it relies entirely on a classical memory, which limits the exploitation of genuinely quantum resources.
By contrast, in our indirect-measurement OLP protocol, although less information is extracted per measurement, the reservoir retains quantum coherences that constitute a form of quantum memory. We therefore expect that, at comparable resource scaling, the optimized OLP can outperform protocols that rely only on classical memory like the feedback-driven protocol proposed in \cite{Kobayashi2024}.

\begin{figure}
    \centering
    \includegraphics[width=0.45\textwidth]{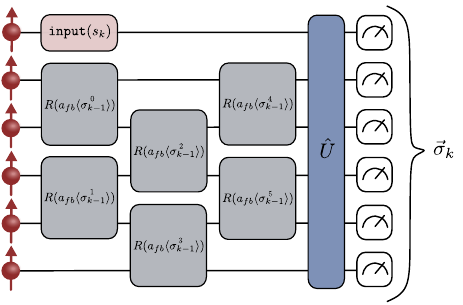}
    \caption{Feedback-driven protocol. We adapt the feedback-driven protocol by inserting the unitary feedback layer in our reservoir pipeline. The measurements after the evolution are projective, and at each new input, we reset the system to the initial conditions.}
    \label{fig:fb_protocol}
\end{figure}

To test this expectation, we adapt the feedback-driven protocol to our spin-reservoir setting and benchmark it directly against our optimized OLP. The adapted protocol operates as follows. For each element \(k\) of the input time series: (i) the value \(s_k\) is encoded into one of the reservoir qubits with the same encoding map of the OLP; (ii) a unitary feedback layer is applied, which re-injects information about the previous expectation values \(\langle \sigma^{i}_{k-1} \rangle\) into the reservoir; (iii) the system is evolved under the reservoir Hamiltonian \eqref{eq:res_ham}; and (iv) all qubits are measured projectively and the reservoir is reset to its initial state before the next input. This routine is repeated sequentially for the entire time-series. To ensure consistency with our previous analysis we measure single-spin observables in all three directions (\(x,y,z\)), but exclude two-spin observables since the classical-feedback protocol was originally devised for single-spin readout.
The feedback layer is implemented as a staggered array of two-qubit modules acting across the reservoir qubits (see Fig.~\ref{fig:fb_protocol}). Each module is parameterized by the measured expectation values at the previous step and a feedback strength $a_{fb}$. This module acting on qubits \(i\) and \(j\) takes the form
\begin{equation}
    R_{ij}(\theta) = RX_i(\theta)RX_j(\theta)CX_{ij}RZ_j(\theta)CX_{ij},
\end{equation}
where \(\mathrm{RX}(\cdot)\) and \(\mathrm{RZ}(\cdot)\) denote single-qubit rotations, \(\mathrm{CX}_{ij}\) is a CNOT with control \(i\) and target \(j\) and $\theta = a_{fb}\langle \sigma_{k-1}^i\rangle$. 

\begin{figure}
    \centering
    \includegraphics[width=0.48\textwidth]{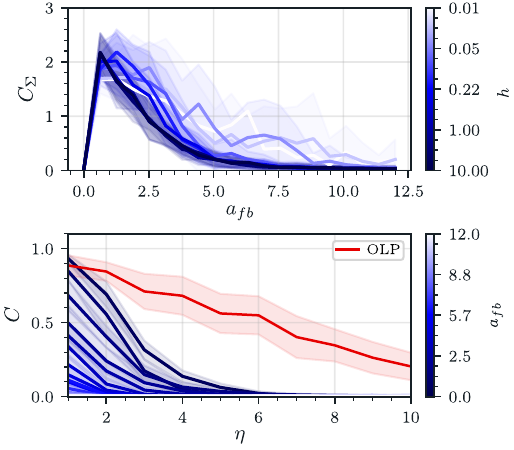}
    \caption{Performance of the feedback-driven protocol across different systems (top) and versus the optimal OLP (bottom) for the short-term memory task. We consider an infinite resource scenario and each line is the average over 10 different realizations. The shaded regions denote one standard deviation around the mean. In the bottom panel, the system chosen for the feedback protocol has an $h$ value of 10.}
    \label{fig:fb_results}
\end{figure}

We test this adapted feedback-driven model on the same short-term memory task considered above. As a first step, we vary the Hamiltonian parameter \(h\), which controls the external field, and evaluate the sum capacity at different feedback strengths $a_{fb}$. The results (Fig.~\ref{fig:fb_results} (a)) show that the maximum sum capacity is obtained for \(h=10\) and $a_{fb}$ $\approx 0.63$.
We then fix \(h=10\) and examine in detail the capacities of individual memory tasks at different delays \(\eta\), while still scanning over $a_{fb}$. These are compared with the corresponding capacities of the optimally performing OLP (Fig.~\ref{fig:fb_results} (b)). The comparison reveals that at short delays (\(\eta=1\)) the optimal feedback-driven protocol performs comparably to the optimized OLP. However, as the task becomes more demanding (larger \(\eta\)), the advantage of the OLP becomes evident: the presence of quantum memory allows it to maintain higher performance, clearly surpassing the classical-memory-based feedback strategy.

\section{Conclusions and Discussion}
In this work, we demonstrate that back-action, a purely quantum effect absent in classical systems, can be exploited as a computational resource, leading to enhanced performance of quantum reservoir computing in temporal data processing tasks. 

By studying the performance ratio $P_R$, we showed that a suitable choice of the measurement strength and external field enables the OLP to match or even surpass the performance of unperturbed-dynamics protocols. In particular, large regions of the $(g,h)$ parameter space display a clear advantage for the OLP, with the strongest enhancements appearing precisely where the unperturbed dynamics is most constrained.
To further characterize this effect, we introduced the stricter measure $P_R^{\min}$, which compares the OLP with the best-performing unperturbed reservoir. This allowed us to isolate the contribution of back-action independently of parameter choice. Notably, even for OLP configurations that already achieve high capacities, we observe that back-action continues to provide a measurable improvement, demonstrating that its beneficial role extends across the relevant operating regimes.

We then extended the analysis to finite-resource scenarios, where both sampling noise and protocol efficiency become relevant. When comparing the OLP with the RSP, the observed enhancements combine the intrinsic advantage provided by back-action with the superior scaling properties of the OLP. To disentangle these effects, we next compared the OLP with the RWP, which shares the same linear scaling in resources. By examining the maximal value of $P_R$ with respect to the RWP, we find that the positive impact of back-action remains significant even at finite sampling and converges toward the ideal infinite-resource behaviour as the number of shots increases. This confirms that the advantage observed for the OLP is a genuine effect of back-action rather than a consequence of differing resource requirements.

Across several sub-tasks derived from two standard benchmark problems, we also highlighted the importance of tuning the OLP parameters. The optimal measurement strength and external fields depend on the specific computational goal, and choosing these parameters appropriately is crucial to unlocking the benefits of back-action.

Finally, by benchmarking the OLP against the feedback-driven protocol, a linear-scaling classical–quantum hybrid scheme, we demonstrated that the OLP’s ability to partially preserve coherences in the reservoir state provides an advantage inaccessible to purely classical feedback mechanisms. Taken together, these findings establish that measurement back-action itself serves as a genuine computational resource for quantum reservoir computing.

As the field transitions to experimental implementations, our results highlight the importance of identifying protocols that make the best use of limited resources while exploiting the beneficial role of measurement back-action. In this context, we combine the previous knowledge on the better scaling of the OLP compared to the RSP with the ability to positively affect the reservoir dynamics by tuning the measurement strength. Together, these findings emphasize the OLP as a compelling candidate for near-term experimental demonstrations of QRC. This practical orientation is essential as experimental interest in QRC continues to grow \cite{Yelin2024, Chen2020, 
Suzuki2022, spagnolo2022experimental,Pfeffer2022, Kubota2023, Yasuda2023, hu2024overcoming, Paternostro2024, PhysRevApplied.20.014051}. From an experimental perspective, the OLP is feasible with currently available technology. Its main requirement is the ability to implement weak or partial measurements with a tunable strength, a capability that has already been demonstrated in several quantum platforms, including superconducting circuits, cold atoms, and photonic systems \cite{pan2020weak, spagnolo2022experimental, murch2013observing, naghiloo2019introduction, Yasuda2023}. In integrated photonics for example, such measurements can be naturally realized by using parametrized beamsplitters or tunable directional couplers that route the signal mode into an ancillary path only with a controlled probability. By detecting the ancilla mode while leaving the main mode largely undisturbed, these devices effectively implement a weak measurement whose strength is set by the beamsplitter transmissivity. This approach, together with the ability to incorporate additional ancilla modes on-chip, provides a practical route for realizing indirect-measurement-based protocols such as the OLP on scalable photonic hardware.

Moreover, our work addresses the growing concerns regarding the scalability and trainability of quantum machine learning models. As variational quantum algorithms face challenges related to these issues, QRC emerges as a compelling alternative. Not only by following this approach trainability issues such as barren plateaus \cite{LaroccaBarrenPlateaus} are avoided. Using the optimized OLP we provide a pathway for conducting machine learning tasks efficiently at scale.

In conclusion, our analysis underscores the fundamental role of indirect measurements in enhancing the performance of QRC through the OLP framework. By demonstrating that making use of indirect measurements not only improves the scaling of the algorithm but can also improve its performance given an appropriate parameter selection, we lay the groundwork for future research.

Looking forward, we recognize the need for extended studies across a broader range of tasks, classical and quantum \cite{Nokkala2023}, and metrics, which may reveal even more generalized conclusions regarding the behavior of QRC algorithms under the OLP. In this direction, studying the effects of evolution time and coherence dynamics may provide deeper insights into the quantum processes underlying these systems. Our results strongly support the experimental realization of QRC algorithm with indirect measurements through the OLP, particularly within emerging platforms that utilize neutral atoms and photons. As quantum computing evolves, these insights will drive both theoretical and practical applications in quantum machine learning.
\section*{Data and code availability statement}

A tutorial version of our implementation of the QRC algorithm utilized in this work, along with the necessary data to partially reproduce some of the figures, is available at \cite{git_tuto}. The complete dataset is available at \cite{data3058_2026}.

\section*{Acknowledgements}
We would like to express our gratitude to Rodrigo Martínez-Peña for their attentive reading and insightful comments on this manuscript.
% To IFISc people:
P.M. also thanks Rodrigo Martínez-Peña, Gian Luca Giorgi, Miguel C. Soriano, and Roberta Zambrini for their stimulating discussions at IFISC in Mallorca, which contributed significantly to the success of this project.
% To Innsbruck people
G.F. would also like to thank Roberto Tricarico and Lisa Bombieri for the careful reading of the manuscript and the fruitful discussions.
G.F. acknowledges support from a ”la Caixa” Foundation (ID 100010434) fellowship. The fellowship code is
LCF/BQ/DI23/11990070.
%for Pere:
This project has received funding from MICIN, the European Union, NextGenerationEU (PRTR-C17.I1), the Government of Spain (Severo Ochoa CEX2019-000910-S and FUNQIP), the ERC AdG CERQUTE and the AXA Chair in Quantum Information Science, the ERC AdG NOQIA, MCIN/AEI (PGC2018-0910.13039/501100011033, CEX2019-000910-S/10.13039/501100011033, Plan National STAMEENA PID2022-139099NB, the “European Union NextGenerationEU/PRTR" (PRTR-C17.I1), FPI); Ministry for Digital Transformation and of Civil Service of the Spanish Government through the QUANTUM ENIA project call - Quantum Spain project, and by the European Union through the Recovery, Transformation and Resilience Plan - NextGenerationEU within the framework of the Digital Spain 2026 Agenda; CEX2024-001490-S [MICIU/AEI/10.13039/501100011033]; Fundació Cellex; Fundació Mir-Puig; Generalitat de Catalunya (European Social Fund FEDER and CERCA program), Barcelona Supercomputing Center MareNostrum (FI-2023-3-0024);  Funded by the European Union (HORIZON-CL4-2022-QUANTUM-02-SGA  PASQuanS2.1, 101113690, EU Horizon 2020 FET-OPEN OPTOlogic, Grant No 899794, QU-ATTO, 101168268),  EU Horizon Europe Program (No 101080086 NeQSTGrant Agreement 101080086 — NeQST).

\section*{Appendix}
\subsection{Details on the finite resource scenario}
\label{appendix_A}
Here we consider the case in which each expectation value extracted from the reservoir is computed with a finite number of shots $N_s$. To take this into account, it has been shown that it is enough to add a Gaussian noise dependent on $N_s$ and $g$ to the previously computed expectation values \cite{mujal2023time} (for further details, refer to Equation (\ref{eqsuncertainties})).

Within this scenario, it is also important to consider the inherent inefficiency of the RSP compared to the OLP as discussed in \cite{mujal2023time}. This analysis builds upon taking into account the time resources that each protocol requires to be executed. 
Let's start by assuming that the time required to reset the reservoir and to take a measurement is negligible. Hence, we can define the experimental time required to run the QRC algorithm with the RSP on a dataset of $K$ elements as:
\begin{equation}\label{eq:t_RSP}
    T_\text{exp} ^\text{RSP} = 3 N_s ^\text{RSP} \left( K' \tau _{wo} + \frac{1}{2} (K'+1)K'\Delta t \right),
\end{equation}  
where $K'$ is equal to $K$ minus the number of samples used for the washout stage $K_{wo}$ and $\tau _{wo} = K_{wo} \Delta t $ is the time required for the washout stage. On the other hand, in the case of using the OLP we get a linear dependence on $K'$ rather than a quadratic:
\begin{equation}\label{eq:t_OLP}
   T_\text{exp} ^\text{OLP} = 3 N_s ^\text{OLP}  \left( K\Delta t \right).
\end{equation}

In both \eqref{eq:t_RSP}, \eqref{eq:t_OLP} the factor $3$ is since we cannot measure simultaneously in all the directions $x$, $y$ and $z$. 

Then, if we fix $N_s$ for the OLP, we expect to have a smaller amount of resources for the RSP given the assumption that we have the same experimental time at our disposal for both protocols. To see this, let's formally set $T = T_\text{exp} ^\text{OLP} = T_\text{exp} ^\text{RSP}$ as the available resource in time and fix the number of shots for the OLP. The available shots for the RSP result to be:
\begin{equation} \label{eq:relNRSPNOLP}
    N_s ^\text{RSP} = N_s ^\text{OLP} \frac{2K}{K'^2+K'(2K_{wo}+1)}.
\end{equation}

\subsection{Independent parameter optimization per direction}
\label{app:indep_dir}
In this appendix we explore the effect of maximizing the minimum performance ratio $P_R^{\min}$ independently for each observable direction. We do this analysis for the short-memory task.  Concretely, we consider three separate scenarios in which only the observables along a single axis (either $x$, $y$, or $z$) are used to compute $P_R^{\min}$, and we determine the optimal parameters $(g,h)$ that maximize $P_R^{\min}$ for that direction following the same procedure described in the main text. This per-direction analysis give intuition about the contribution of each direction to the total performance.

\begin{figure*}
    \centering
    \includegraphics[width=\textwidth]{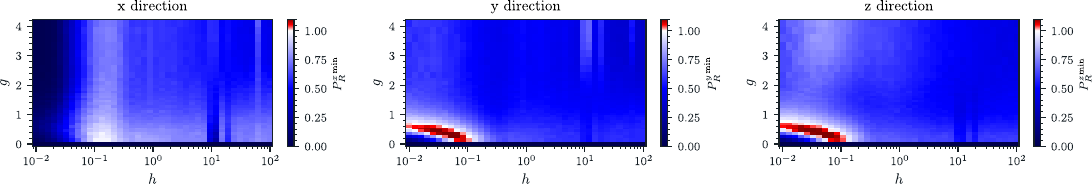}
    \caption{Minimum performance ratio analysis for the short-term memory task, computed independently for each measurement direction. Each panel shows the $g-h$ map when only observables in a single direction are used to evaluate $P_R^{\min}$. The optimal parameters and corresponding $P_R^{\min}$ values for each direction are: ($x$). RSP: $h'^* = 0.170$; OLP : $g^* = 0.150$, $h^* = 0.170$, $P_R^{\min} = 0.981$ ($y$). RSP: $h'^{*}=0.084$; OLP: $g^{*}=0.355$, $h^{*}=0.041$, $P_R^{\min}=1.145$ ($z$). RSP: $h'^{*}=0.084$; OLP: $g^{*}=0.355$, $h^{*}=0.041$, $P_R^{\min}=1.159$.}
    \label{fig:cmaps_each_direction}
\end{figure*}

As shown in Figure \ref{fig:cmaps_each_direction}, the $x$ direction contributes less to $P_R^{\min}$ than the other two directions, which are very similar both in behaviour and in optimal parameters.

\begin{figure}[t!]
    \centering
    \includegraphics[width=0.48\textwidth]{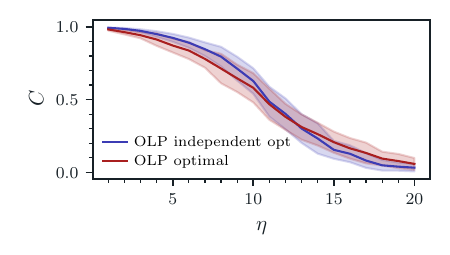}
    \caption{Capacity for different sub-tasks. The figure compares the capacity of the OLP when optimized independently for each measurement direction and combined across directions with the OLP optimized jointly over all observables. The per-direction strategy gives a modest gain, redistributing capacity from long to short delays (see main text for discussion).}
    \label{fig:optimal_capacity_smt_eachdirection}
\end{figure}

Next, we take the observables corresponding to each direction's independently optimized parameters and build a combined dataset used to solve the task. In Figure \ref{fig:optimal_capacity_smt_eachdirection} we compare the capacities of this combined dataset with those obtained from the joint (global) optimization reported in the main text. The combined, per-direction strategy yields a small increase in the performance ratio, $P_R^{\min}=1.040$ (vs.\ $P_R^{\min}=1.029$ for the global optimisation). Although this increase has little impact on the summed capacity reported in the main text, the independent optimisation redistributes capacity from larger delays (the "tail") to shorter delays (the "head"): capacities at small delay $\eta$ are improved while long-$\eta$ performance is slightly reduced. Thus, for applications that prioritise short-delay memory, the per-direction strategy can be advantageous. Finally, since the $x$ direction contributes relatively little, one may reasonably omit $x$ observables when resources are constrained and focus on the more informative $y$ and $z$ directions.
\\

\newpage

\bibliography{biblio}

\end{document}